\documentclass[preprint,12pt]{elsarticle}

\usepackage{amssymb}
\usepackage{amsmath}
\usepackage{graphicx}
\usepackage{subfigure}

\begin{document}

\begin{frontmatter}



\title{Fuel burnup analysis of the TRIGA Mark II Reactor at the University of Pavia}


\author[MIB,INFNMIB]{Davide Chiesa}
\author[MIB,INFNMIB]{Massimiliano Clemenza}
\author[MIB,INFNMIB]{Stefano Pozzi}
\ead{stefano.pozzi@mib.infn.it}
\author[MIB,INFNMIB]{Ezio Previtali}
\author[MIB,INFNMIB]{Monica Sisti}
\author[LENA,INFNPV]{Daniele Alloni}
\author[LENA,INFNPV]{Giovanni Magrotti}
\author[LENA,INFNPV]{Sergio Manera}
\author[LENA,INFNPV]{Michele Prata}
\author[LENA,INFNPV]{Andrea Salvini}
\author[POLI,INFNMIB]{Antonio Cammi}
\author[POLI]{Matteo Zanetti}
\author[SISSA]{Alberto Sartori}

\address[MIB]{Physics Department ``G. Occhialini" of Milano-Bicocca University, piazza della Scienza 3, 20126 Italy}
\address[INFNMIB]{INFN section of Milano-Bicocca, piazza della Scienza 3, 20126 Italy}
\address[LENA]{Laboratorio Energia Nucleare Applicata (L.E.N.A.) of the University of Pavia, via Aselli 41, Pavia 27100, Italy}
\address[INFNPV]{INFN section of Pavia, via A. Bassi 6, Pavia 27100, Italy}
\address[POLI]{Politecnico di Milano, Department of Energy, CeSNEF (Enrico Fermi Center for Nuclear Studies), via La Masa 34, 20156 Milano, Italy}
\address[SISSA]{SISSA mathLab, International School for Advanced Studies,\\Via Bonomea 265, 34136 Trieste, Italy}

\begin{abstract}
A time evolution model was developed to study fuel burnup for the TRIGA Mark II reactor at the University of Pavia. The results were used to predict the effects of a complete core reconfiguration and the accuracy of this prediction was tested experimentally. We used the Monte Carlo code MCNP5 to reproduce system neutronics in different operating conditions and to analyse neutron fluxes in the reactor core. The software that took care of time evolution, completely designed in-house, used the neutron fluxes obtained by MCNP5 to evaluate fuel consumption. This software was developed specifically to keep into account some features that differentiate experimental reactors from power ones, such as the daily \textsc{on/off} cycle and the long fuel lifetime. These effects can not be neglected to properly account for neutron poison accumulation. We evaluated the effect of 48 years of reactor operation and predicted a possible new configuration for the reactor core: the objective was to remove some of the fuel elements from the core and to obtain a substantial increase in the Core Excess reactivity value. The evaluation of fuel burnup and the reconfiguration results are presented in this paper.

\end{abstract}


\end{frontmatter}


\section{Introduction}
\label{Introduction}

The TRIGA Mark II Reactor at the Applied Nuclear Energy Laboratory (L.E.N.A.) of the University of Pavia is a pool type reactor with a nominal  power of 250 kW. It was brought to its first criticality in 1965 and since then it was used for several scientific activities, such as radioisotope production, material analysis via neutron activation and reactor physics studies. \\
The reactor core is shaped as a right cylinder featuring 90 slots, distributed over 5 concentric rings, which can contain either fuel elements, graphite (dummy) elements, control rods or irradiation channels. The fuel consists of a uniform mixture of uranium (8\% wt., enriched 20\% wt. in $^{235}$U), zirconium (91\% wt.) and hydrogen (1\% wt.). The Fuel Elements (FEs) used in the current core configuration belong to different manufacturing series that were designed by General Atomics over the years. The 101-type FEs are characterized by aluminum cladding, 1:1 atomic ratio between zirconium and hydrogen and two burnable poison disks containing samarium. The 103-type and 104-type FEs have stainless steel cladding, 1:1.6 Zr–H ratio and a zirconium rod at the center of the fuel; furthermore, a burnable poison disk containing molybdenum is present in the 104-type FEs.

\begin{figure}[h!]
	\begin{center}
		\includegraphics[width=0.5\textwidth]{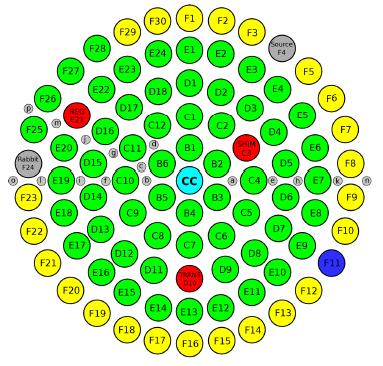}
		\caption{Core configuration in 1965. Fuel rods are represented in green, graphite rods in yellow, control rods in red, irradiation channels in gray and an empty channel in blue. CC = central irradiation channel (or central thimble).}
		\label{Fig_Core}
	\end{center}
\end{figure}

In the recent years, the TRIGA Mark II reactor of Pavia was characterized in detail by means of both neutron activation measurements in different irradiation facilities \cite{Flussi}\cite{Bayes}\cite{Bayes2}\cite{AlCo} and development of proper simulation tools for modeling the neutronics, the dynamics and the thermal-hydraulics of the system \cite{Tesi}\cite{Borio}\cite{CaldoPulito}\cite{Cammi}\cite{Sartori}.\\
In particular, to study the reactor neutronics, we developed a Monte Carlo model based on MCNP5 \cite{MCNP}. We chose this code thanks to its general geometry modeling capabilities, correct representation of neutron transport effects and continuous-energy cross section treatment. This model was extensively tested and validated by reproducing the experimental results obtained in 1965, during the operations that followed the reactor first startup \cite{Cingoli}. This analysis included the reproduction of control rod calibration curves and the evaluation of the criticality coefficient ($k_{eff}$) in several reactor configurations both at low power ($\sim 10$ W) and full power (250 kW) conditions~\cite{Tesi}\cite{CaldoPulito}.\\
In this initial configuration (Fig. \ref{Fig_Core}) all the fuel elements were new and did not contain significant amounts of neutron poisons. Therefore, in the MCNP5 simulations, we could model the fuel using the data of the original isotopic composition provided by the manufacturer.\\
In this work, through the analysis of the fuel burnup, we aim to characterize the TRIGA reactor in its present configuration, after many years of operation. Moreover, this analysis is important to study the fuel cycle and determine the amount of long-lived radioactive waste which are produced in a reactor after a certain operating time.

\section{Reactor fuel evolution}

In order to characterize and model the current configuration of the TRIGA Mark II reactor, we must take into account several aspects related to fuel burnup. First of all, the consumption of the fissile isotopes and the buildup of neutron poisons affects the system reactivity. Therefore, in order to simulate the criticality condition of the reactor, it is crucial to know the fuel composition as function of time. Moreover, since the fuel burnup is not homogeneous within the core, also the neutron fluxes and the power distributions change over time. In addition, we must consider that, during the operations of refueling and core reconfiguration, some fuel elements are replaced, added or moved to different locations.\\
For this reason, to evaluate the fuel composition as function of time, the whole history of the nuclear reactor must be followed, reproducing all the different configurations of the core and collecting the data concerning operating times and powers.\\
In order to analyze and simulate the fuel burnup in the TRIGA Mark II reactor, we decided to develop a time evolution software which couples the historical data of reactor operating time and core configurations with the information about effective cross sections and neutron fluxes, evaluated through the MCNP5 model. This software was completely developed in-house to take into account some features that differentiate the experimental reactors from those used for power production, such as the daily \textsc{on/off} cycle and the long fuel lifetime. These effects are not negligible when properly accounting for neutron poison build-up in this kind of systems.\\
The historical data about the critical configurations of the core can then be used as a benchmark for validating the MCNP5 model after each step of the burnup calculation. In this way, at the end of the simulation process, we aim to obtain a reliable and benchmarked simulation model for the current reactor configuration.

\section{Burnup calculation strategy}
The simulation model for the fuel burnup is based on the solution of a coupled set of differential equations, whose variables are the concentrations of all the isotopes in the fuel \cite{Stacey}. Although all the nuclear reactions and decays should be considered for an exact calculation of the fuel evolution, the main relevant processes to be modelled in order to ensure a good accuracy in the neutronics simulations are fission, neutron capture and radioactive decays.\\
The isotopic concentration $n_{j}(t)$ of a fission product species $j$, characterized by $\lambda_j$ decay constant and $\sigma_a^j$ neutron absorption cross section, evolves in time according to this generic formula \cite{Stacey}:\\

\begin{equation}
	\label{EQ}
	\dfrac{dn_{j}}{dt} = \sum_{k}\gamma_j^{k} \Sigma_f^{k} \Phi+\sum_{i} \left( \lambda^{(i\rightarrow j)}+\sigma^{(i\rightarrow j)}\Phi\right)n_i - \left(\lambda^j+\sigma_a^j\Phi\right)n_j,
\end{equation}

\noindent where the sum $\sum_k$ is performed over all fissionable nuclei, $\gamma_j^k$ is the fission yield of isotope $j$, $\Sigma_f^k$ is the macroscopic fission cross section, $\Phi$ is the integral neutron flux intensity, $\lambda^{(i\rightarrow j)}$ is the decay rate of isotope $i$ to produce isotope $j$ and $\sigma^{(i\rightarrow j)}$ is the transmutation cross section for the production of isotope $j$ by neutron capture in isotope $i$.\\
In order to take into account the reaction rate dependence on the neutron spectrum, the cross sections in Eq. \ref{EQ} must be calculated as effective ones, i.e. average cross sections weighted by the energy distribution of the neutron flux. The time evolution of the elements belonging to the original fuel composition (U, Zr and H) or produced exclusively through neutron capture can still be described by Eq. \ref{EQ} by setting the fission yield value $\gamma_j$ equal to 0.\\
Equation \ref{EQ} must be integrated to determine fuel composition changes over its lifetime. In order to carry out this calculation, however, the time dependence of the neutron flux must be known. To overcome this issue, the 48 year period is divided in several time intervals in which the neutron flux distribution is assumed to vary negligibly. For the TRIGA Mark II reactor, we verified, through dedicated MCNP5 simulations, that the time scale for significant flux changes is greater than the maximum operation time elapsed between two core reconfigurations ($\simeq$3500 hours). Particularly, we evaluated that the neutron flux variations due to the fuel burnup after $\sim3500$ hours at 250kW are less than $3\%$ in all the core positions. On the other hand, we also found that the neutron flux can vary more than $10\%$ after a core reconfiguration. For this reason, we carried out the burnup calculation using 27 time steps, one for each core reconfiguration occurred between 1965 and 2013.\\
The applied approximations are justified because their contribution has a smaller effect than the unavoidable uncertainties related to normal reactor operation (for instance, continuous movement of the control rods). To further prove this statement, in Section \ref{Sec_Bench} we study the variation of $k_{eff}$ over the course of the years, checking if our approximations introduce any kind of unwanted time dependence on the data.\\
Our software evaluates the time evolution of each fuel element by iterating the following procedure over all the time steps:\\

\begin{enumerate}
	\item a MCNP5 simulation generates the neutron flux distribution in the reactor core. This simulation is performed by taking into account the thermal effects arising at full power (250kW) and reproducing the original positions of the control rods \cite{CaldoPulito};
	\item flux spectra and ENDF/B-VII cross section data libraries are combined to evaluate the effective cross section for the isotopes of interest;
	\item data about neutron fluxes, fission yields, cross sections, radioactive decay chains and reactor operating time are combined to calculate the new isotopic composition of the fuel elements, according to Eq.\ref{EQ};
	\item according to the new configuration, a new MCNP5 input file is prepared, where the fuel elements are simulated with the new isotopic composition and repositioned in the reactor core.
\end{enumerate}

To simplify the problem solution, we applied some approximations:

\begin{enumerate}

	\item Each fuel element was divided in 5 axial sections, to account for the uneven distribution of neutron flux over the vertical axis \cite{AlCo}\cite{CaldoPulito}. Configurations with more axial or radial sections were tested, but the results were found to be compatible within the model uncertainties. Therefore, we chose to model 5 sections, thus saving computation time without losing accuracy. 
	\item The list of all the fission products and trans-uranium elements to be included in the MCNP5 simulation was simplified, including only the isotopes impacting on the system neutronics in a non-negligible way (Tab. \ref{Tab_Isotopes}). These isotopes were identified by calculating their absorption cross sections and the maximum concentrations they could reach in the TRIGA reactor during 48 years of operation\footnote{The maximum concentrations were evaluated through an approximate solution of Eq. \ref{EQ}, which was purposely overestimated to be conservative.}. These data were then compared to those of ${}^{135}$Xe, whose impact on $k_{eff}$ is experimentally known, and an isotope was finally neglected if its maximum impact on $k_{eff}$ was evaluated to be less than $0.01\%$.	
	\item Each of the 27 time steps was divided in sub-intervals, to account for the reactor daily \textsc{on/off} cycle. Since an accurate reconstruction of the real cycles was impractical, we chose 6 hour sub-intervals for the \textsc{on} condition and we calculated the duration of the \textsc{off} sub-steps to match the real time difference between the two reconfiguration dates.
	\item The calculation for ${}^{135}$Xe was carried out separately, due to its large daily variations; its average concentration was estimated by considering a typical week in which the reactor operates 6 hours a day, from Monday to Friday.

\end{enumerate}

\begin{table}[h!]
	\begin{center}
		\begin{tabular}{|c|c|c|c|c|c|c|c|c|}
			\hline
			\multicolumn{5}{|c|}{\textbf{Fission Products}} && \textbf{Uranium} && \textbf{\small{Trans-Uranium}}\\
			\hline\hline
			$^{83}$Kr & $^{102}$Ru & $^{131}$Xe & $^{145}$Nd & $^{152}$Sm && ${}^{235}$U && $^{239}$Pu\\
			\hline
			$^{95}$Nb & $^{103}$Rh & $^{133}$Cs & $^{147}$Pm & $^{153}$Sm && ${}^{236}$U && $^{240}$Pu\\
			\hline
			$^{95}$Mo & $^{105}$Rh & $^{135}$Cs & $^{147}$Sm & $^{153}$Eu && ${}^{238}$U && $^{241}$Pu\\
			\hline
			$^{97}$Mo & $^{105}$Pd & $^{139}$La & $^{149}$Sm & $^{155}$Eu && && $^{241}$Am\\
			\hline
			$^{99}$Tc & $^{113}$Cd & $^{141}$Pr & $^{150}$Sm & $^{155}$Gd && && \\
			\hline
			$^{101}$Ru & $^{129}$I & $^{143}$Nd & $^{151}$Sm & $^{157}$Gd && && \\
			\hline
			
		\end{tabular}
		\caption{List of the fission products, uranium trans-uranium isotopes included in the burnup calculation.}
		\label{Tab_Isotopes}
	\end{center}
\end{table}

\section{Fuel Evolution Results}
\begin{figure}[h!]
	\centering
	\subfigure{\includegraphics[width=0.45\textwidth]{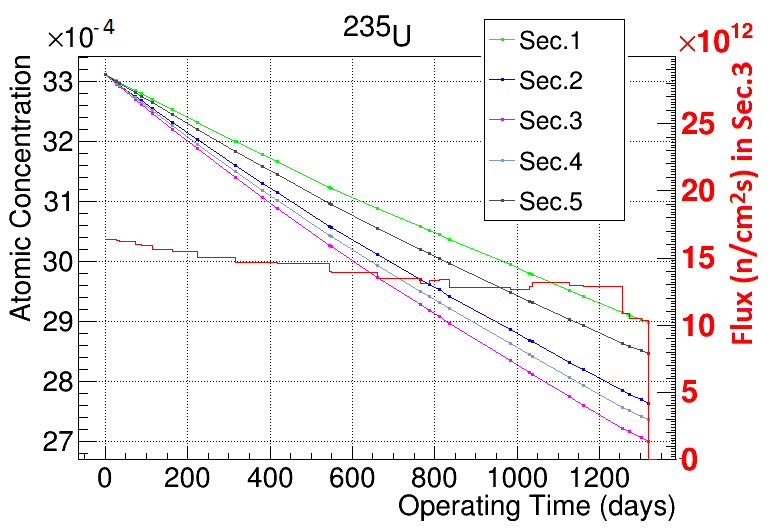}}
	\subfigure{\includegraphics[width=0.45\textwidth]{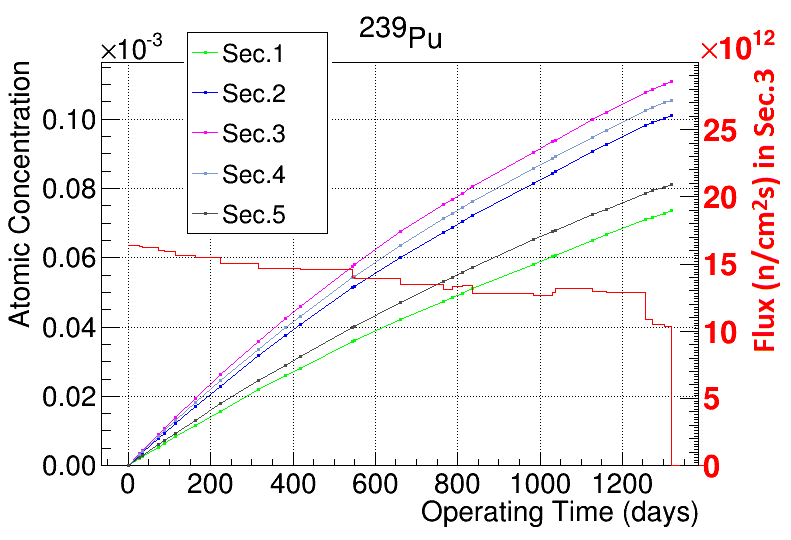}}
	\subfigure{\includegraphics[width=0.45\textwidth]{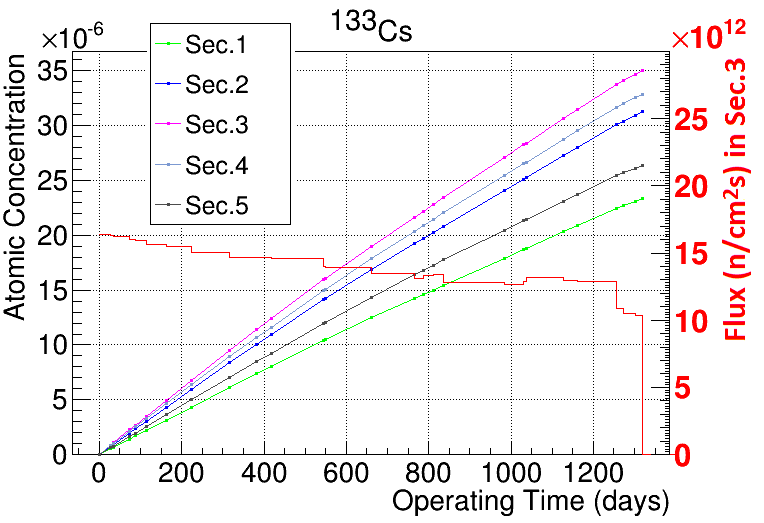}}
	\subfigure{\includegraphics[width=0.45\textwidth]{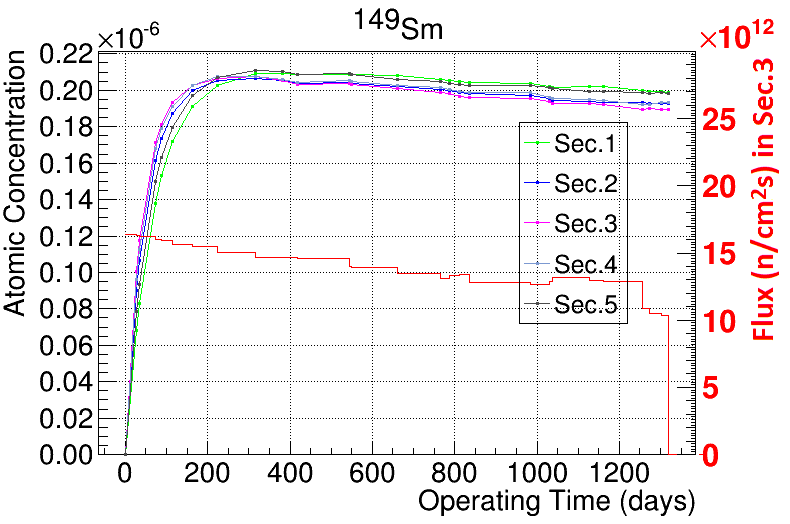}}
	\subfigure{\includegraphics[width=0.45\textwidth]{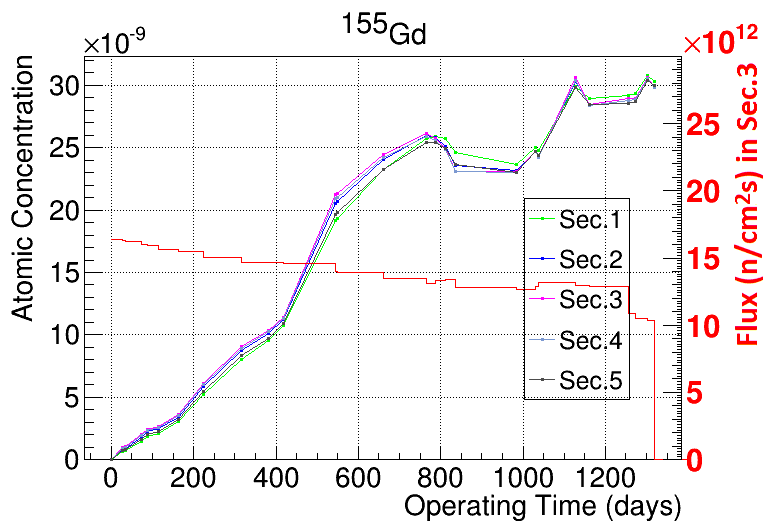}}
	\subfigure{\includegraphics[width=0.45\textwidth]{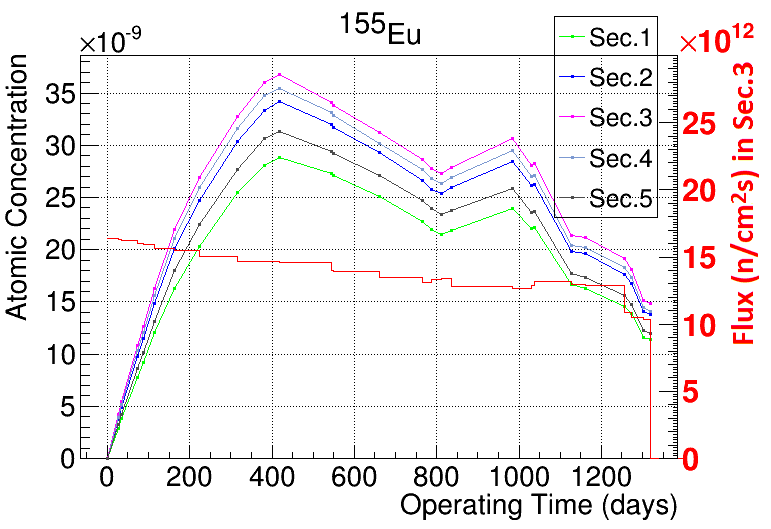}}
	\caption{Time evolution of some atomic concentrations in the 5 sections of a fuel element. The integral neutron flux in the central section is represented in red. The horizontal axis represents the net reactor operating time at full power.}
	\label{Plots}
\end{figure}

In order to analyze the burnup calculation results, we plotted the concentration of the different isotopes as a function of the total reactor operating time. As an example, in Fig. \ref{Plots}, we show the evolution of some isotopes in a specific fuel element, which was used in the TRIGA reactor from 1965 to 2013, changing sometimes its position in the core. The isotopic concentrations in the 5 axial sections of this FE are plotted versus the net operating time of the reactor. In addition, the average neutron flux in the central section (labeled section 3 in Fig. \ref{Plots}) of the fuel element is reported in red.

In Fig. \ref{Plots} we chose to show a set of representative isotopes, characterized by different, peculiar behaviours.

\begin{itemize}
	\item The fissile isotopes ${}^{235}$U and ${}^{239}$Pu. While the first one is burned and its concentration decreases in time, ${}^{239}$Pu is steadily produced by capture reactions on ${}^{238}$U.
	\item $^{133}$Cs, a fission product, which does not reach saturation and follows an approximately linear time profile;
	\item $^{149}$Sm, a fission product, characterized by a high absorption cross section, which reaches saturation; the saturation level changes over time due to the variations of the neutron flux and due to the different concentrations of the fissile elements.
	\item $^{155}$Eu and $^{155}$Gd, whose evolution is correlated, because $^{155}$Gd is produced by the decay of $^{155}$Eu (4.75y of half life).
\end{itemize}

The case of $^{155}$Eu and $^{155}$Gd is particularly interesting: the relatively short half life of ${}^{155}$Eu makes the concentration of those two isotopes strongly dependent on reactor \textsc{off} time (more than any other we considered), during which the decay continues to occur. Since the ratio between \textsc{on} and \textsc{off} time varies over the years, $^{155}$Eu concentration does not reach a stable value. Without the inclusion of the \textsc{off} time evaluation in our model, it would have been impossible to correctly calculate the concentration of these isotopes and, therefore, the final $k_{eff}$ value would have been incorrect.

\section{Benchmark analysis of the fuel evolution model}
\label{Sec_Bench}

\begin{figure}[h!]
	\begin{center}
		\includegraphics[width=\textwidth]{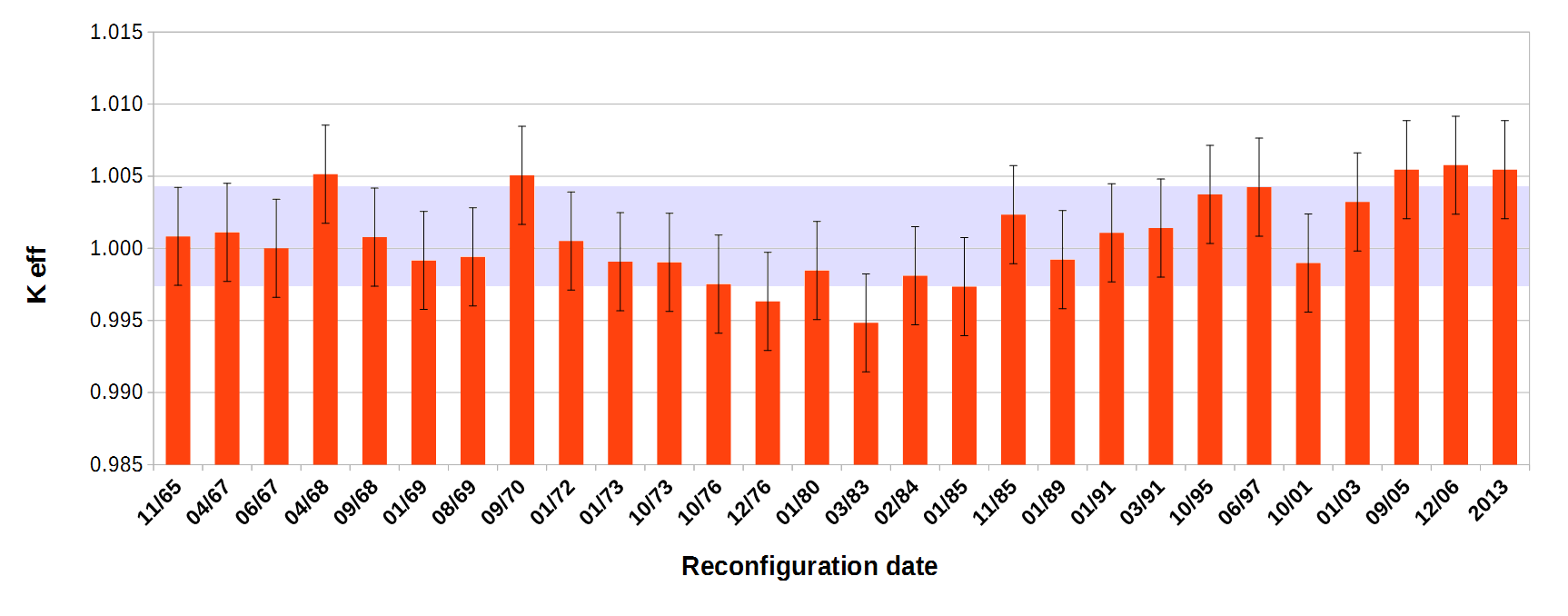}
		\caption{$k_{eff}$ values obtained by the Monte Carlo simulations at each time step. The shadowed area corresponds to the $\pm 1\sigma$ range from the average value, $\sigma$ being the standard deviation of the results. The reported error comes from the statistical and systematic components.}
		\label{Reconf}
	\end{center}
\end{figure}

In order to check the MCNP5 model response to the fuel composition changes, we analysed the variation of the $k_{eff}$ value over the years, following each core reconfiguration. For each simulation the control rods were positioned as to reproduce the first full power criticality following the reconfiguration. As shown in Fig.\ref{Reconf}, all the results are within $\pm 0.005$ with respect to the expected value, $k_{eff}=1$. The systematic uncertainty associated to each $k_{eff}$ value is obtained by summing two contributions: the uncertainty related to 
material composition in the model of the original reactor configuration at low-power~\cite{Tesi}, equal to $\sim0.0018$, and the one associated with the simulation of the thermal effects at full-power~\cite{CaldoPulito}, equal to $\sim0.0015$.

By calculating the average value and the standard deviation of the results, we obtain $k_{eff}=1.00067\pm 0.00291$, which is in good agreement with the expected value. 
Taking into account the fact that this calculation was performed over a period of 48 years, the analysis confirms that the fuel aging was simulated with good accuracy. In fact, we do not observe any particular trend in the results, which could have been linked to an over- or an under-estimation of fuel aging. The fluctuations that affect our data are most probably due to imprecise informations in the historical data.\\

\begin{figure}[h!]
	\begin{center}
		\includegraphics[width=1\textwidth]{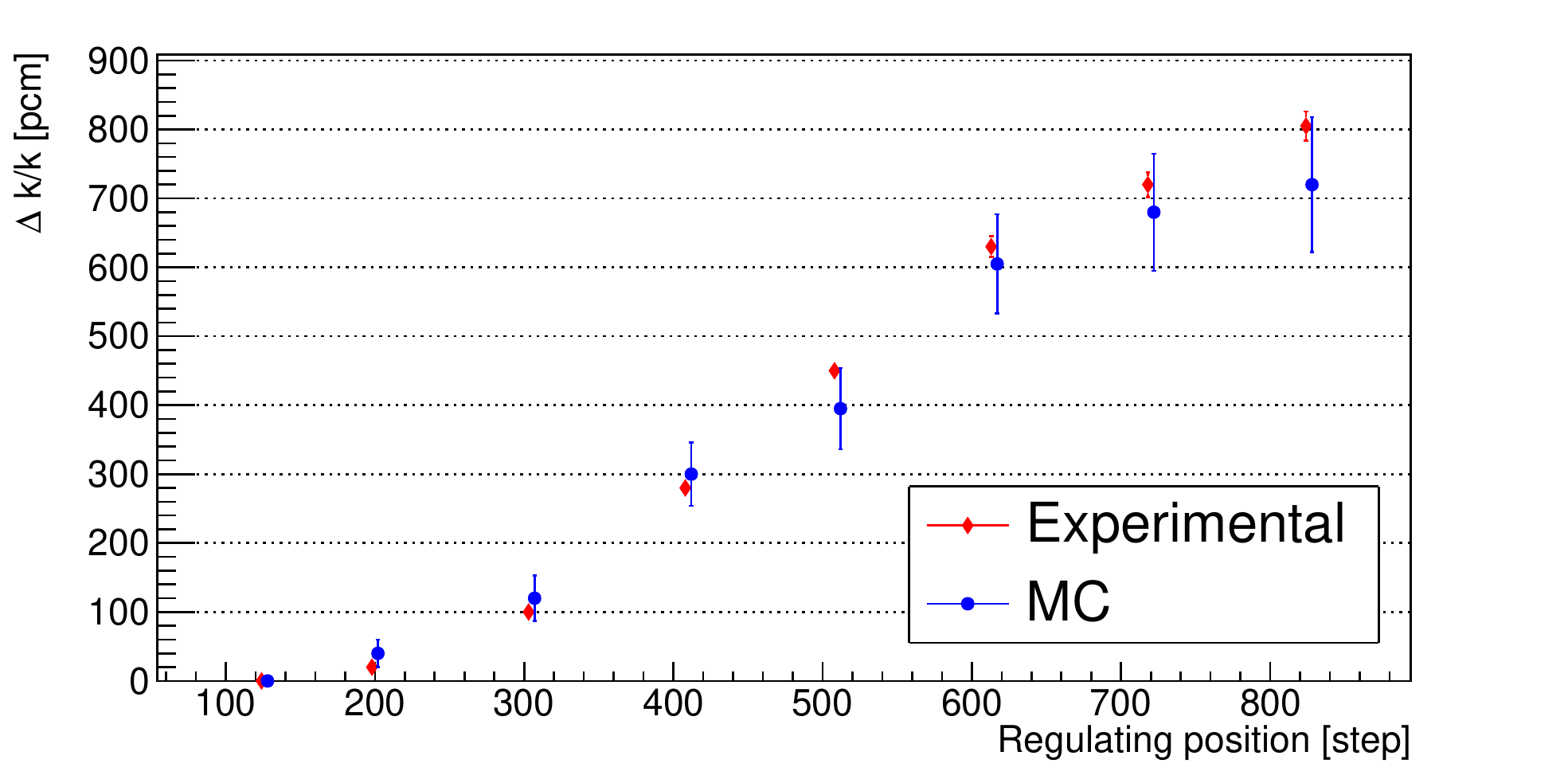}
		\caption{Comparison between experimental and Monte Carlo calibration curve for the Regulating control rod, referring to the July 2013 configuration. The error bars are associated to the statistical component only.}
		\label{REG}
	\end{center}
\end{figure}

As a further benchmark, we checked the ability of the Monte Carlo model to reproduce the system reactivity at low power. In this way, the thermal effects that greatly affect the full power evaluation of $k_{eff}$ can be neglected. The 2013 configuration updated with the computed composition was used. At first, a simulation was performed with the control rods in the positions recorded during a low power $(\sim 1.5W)$ measurement dated September 9, 2013. The result we obtained is

\begin{equation*}
	k_{eff} = 1.00117\pm 0.00025(stat)\pm 0.00180(syst),
\end{equation*}

\noindent as opposed to the expected value of 1. In this case, the systematic component comes from the uncertainty related to material composition in the model of the original reactor configuration at low-power.\\
As a second test, we reproduced the Regulating control rod calibration curve, obtained experimentally in July 2013. The results of the test are shown in Fig. \ref{REG}. To conform with the experimental data format, the control rod positions are expressed in steps (1 step $\simeq 0.05$~cm displacement from the fully inserted position) and the reactivity $\rho$ in \textit{pcm} (\textit{per cent mille}):

\begin{equation*}
	\rho = \dfrac{(k_{eff}-1)}{k_{eff}}
\end{equation*}
\begin{equation*}
	1~pcm = 10^{-5}.
\end{equation*}

Both tests show a very good agreement between experimental results and Monte Carlo simulations: the calibration curve is reproduced with great accuracy and the low power simulation is compatible with the expected values, considering the systematic error component.\\
Finally, as a further demonstration of the reliability of the simulation model, a benchmark analysis concerning the neutron fluxes is presented in \cite{AlCo}, where the  measured data of the integral and fast fluxes, collected in a experimental campaign dated 2013, were compared with those evaluated through the corresponding MCNP5 model, obtaining a good agreement.

\subsection{Core reconfiguration}

The simulation model developed for the 2013 configuration of the TRIGA reactor was then exploited to identify a new optimized core configuration, with the aim of increasing system reactivity while decreasing the total number of fuel elements in the system.\\
An important parameter to be considered for this kind of analysis is the Core Excess (CE), defined as the system reactivity which would be obtained if all the control rods were completely withdrawn from the core. The Core Excess value must be high enough to compensate for the reactivity losses due to thermal effects and $^{135}$Xe poisoning.\\
On September 9, 2013, the Core Excess was evaluated both at low ($\sim10$ W) and full ($250$ kW) power; the measurement was performed on a Monday, to avoid the influence of the ${}^{135}$Xe accumulated during the previous week. The results we obtained were CE${}_{low}=1504\pm18$~pcm and CE$_{high}=314\pm18$~pcm for the low-power and full-power configurations, respectively; these values were particularly alarming, since the $^{135}$Xe accumulated in just one day of operation would be enough to make the reactor inoperable at full power the subsequent morning. To solve this issue, a core reconfiguration was needed in order to increase the reactor Core Excess by at least $300$~pcm. For this purpose, we exploited the MCNP5 model with the fuel burnup calculation extended up to September 2013. New configurations, ensuring a higher Core Excess and requiring less fuel elements in the reactor core, were simulated and analyzed.\\
The general idea behind the choice of those configurations was to put the fuel elements with the higher content of fissile material in the innermost region of the core, where the neutron flux is higher. We ranked the fuel elements from best to worst using a Burnup Index (B.I.), which we defined as follows:

\begin{equation}
	B.I.(t) = \dfrac{1}{n_{235}(0)}\left[n_{235}(0)-\left(n_{235}(t)+n_{239}(t)\dfrac{\sigma_{239}^{f}}{\sigma_{235}^{f}}\right)\right];
\end{equation}

\begin{figure}[h!]

	\begin{center}
		\subfigure{\includegraphics[width=0.45\textwidth]{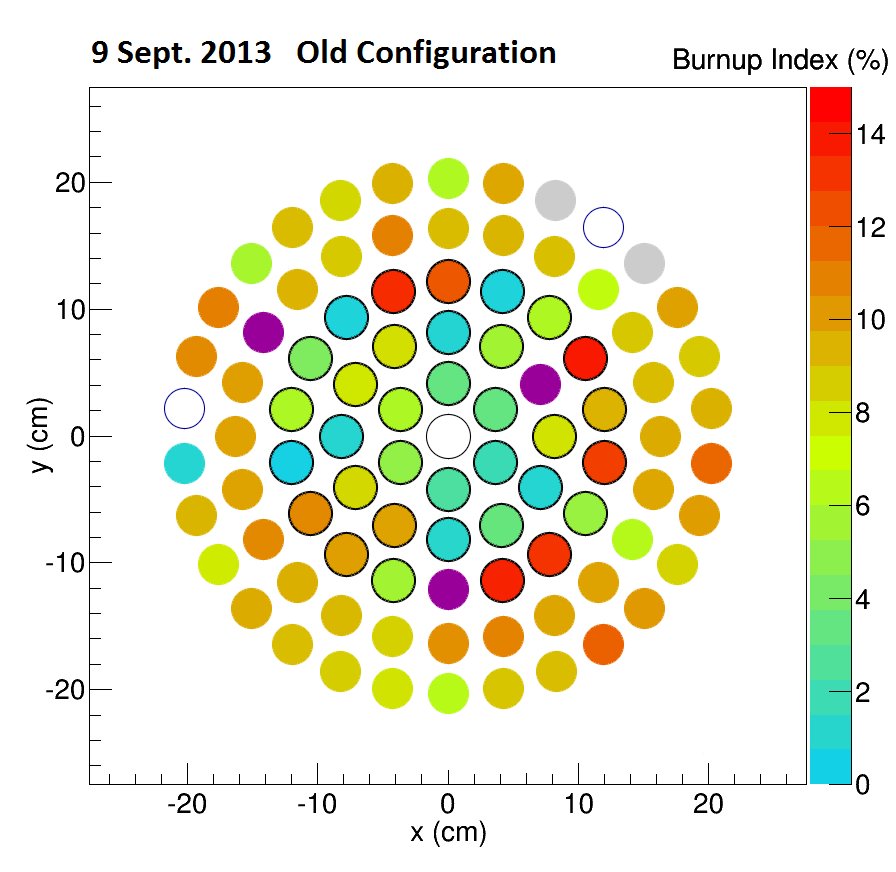}}
		\subfigure{\includegraphics[width=0.45\textwidth]{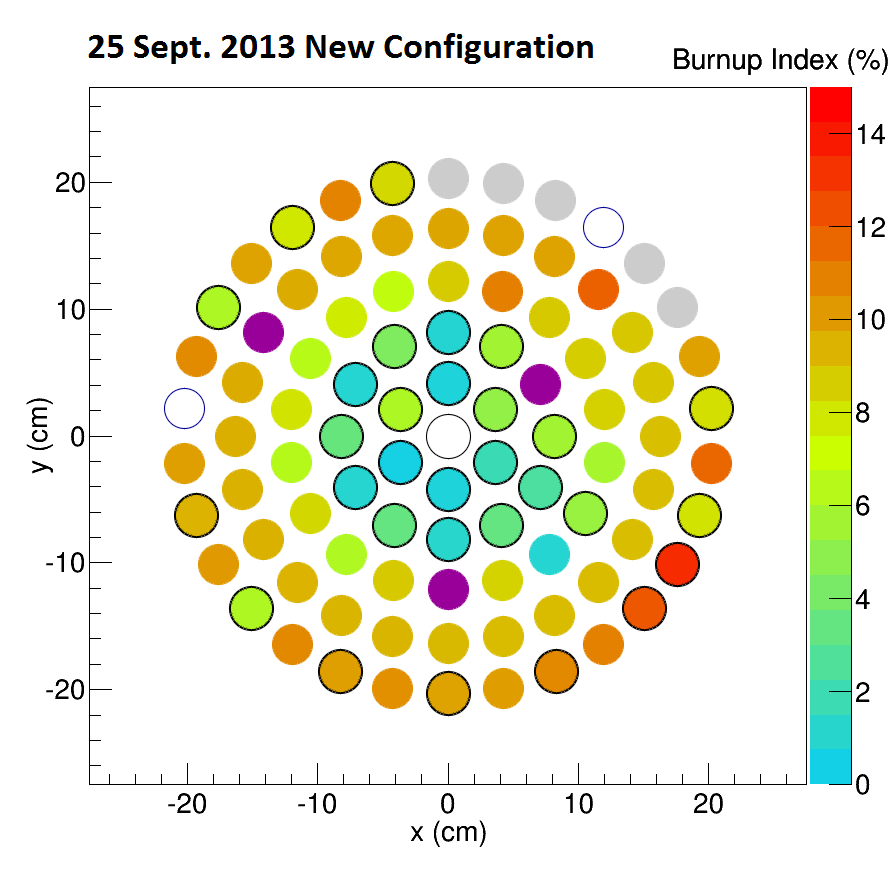}}
		\caption{Comparison between the old and new core configurations. Control rods are in purple, graphite rods in gray. Fuel elements with a black border have a stainless steel cladding, while the others have an aluminium one.}
		\label{OldNew}
	\end{center}
	
\end{figure}

\begin{figure}[h!]
	\begin{minipage}{0.5\textwidth}
		\includegraphics[width=0.95\textwidth]{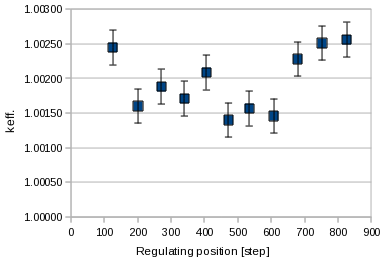}
	\end{minipage}
	\begin{minipage}{0.5\textwidth}
		\begin{tabular}{|c|c|c|}
			\hline
			\textbf{Reg. $[$step$]$} & \textbf{Shim $[$step$]$} & \textbf{k$_{eff}$}\\
			\hline
			126 & 477 & 1.00245\\
			201 & 460 & 1.00160\\
			271 & 447 & 1.00188\\
			340 & 432 & 1.00171\\
			406 & 406 & 1.00208\\
			472 & 379 & 1.00140\\
			534 & 353 & 1.00157\\
			608 & 313 & 1.00146\\
			679 & 295 & 1.00228\\
			752 & 278 & 1.00251\\
			826 & 262 & 1.00256\\
			\hline
		\end{tabular}
	\end{minipage}
	\caption{Set of critical reactor configuration, obtained after the core reconfiguration. The (statistical) error on $k_{eff}$ is equal to $\pm 0.00025$.}
	\label{Crit}
\end{figure}

\noindent $\sigma_{235}^{f}$ is the fission cross section of $^{235}$U, $\sigma_{239}^{f}$ the fission cross section of $^{239}$Pu, $n_{235}(t)$ and $n_{239}(t)$ the concentrations of $^{235}$U and $^{239}$Pu at time t. With this index, we keep into account not only the consumption of the original fissile isotope, but also the accumulation of $^{239}$Pu over time, whose effect has been determined to be significant.\\
The new core configuration, chosen to comply with reactor safety requests and technical prescriptions, contains 80 fuel elements, as opposed to the 83 elements in the 9 September 2013 configuration. The comparison between the two configurations is reported in Fig.\ref{OldNew}, along with the Burnup Index scale.\\
The reactivity gain predicted by the MCNP5 simulations was equal to $416\pm37$~pcm, corresponding to a Core Excess value at low power of $1920\pm37$~pcm. Right after the reconfiguration, that took place on 25 September 2013, an experimental measurement of the CE was performed, in order to check the outcome of the procedure. The new experimental CE resulted equal to $1851\pm22$~pcm, which is a value compatible within $2\sigma$ of the CE predicted by our software. The final reactivity increase, equal to $325\pm22$~pcm, was completely satisfactory and proved that the model had successfully reproduced the fuel evolution over the reactor lifetime.\\
To further verify the effectiveness of our model, we simulated a set of critical reactor configurations, obtained during the operations that followed the reconfiguration. In all these configurations the expected $k_{eff}$ value is $1$. Simulation results are shown in Fig. \ref{Crit}.

\noindent
The average $k_{eff}$ value obtained by the simulations was
\begin{center}
	$k_{eff}=1.00196\pm 0.00043 (stat) \pm 0.00180 (syst)$,\\
\end{center}
which is in full agreement with the expected value $k_{eff}=1$.

\section{Conclusions}

We developed an improved MCNP5 model for the TRIGA Mark~II reactor in which the effects of fuel aging have been simulated.\\
The effectiveness of the time evolution model was tested in multiple ways. The time profile of the $k_{eff}$ value over the 48 years of operation shows no particular trend, indicating that no significant overestimation or underestimation of fuel burnup took place; the average $k_{eff}$ value over this time period, equal to $1.00067\pm 0.00291$, is compatible with the expected value of 1. The Regulating control rod calibration curve, measured on July 2013, was reproduced completely and with good accuracy. A low power system configuration, dated 9 September 2013, was simulated and led to results that, considering the systematic errors, are compatible with the experimental data.\\
Once the effectiveness of the simulation model was proven, its results were used as a tool to predict the effect of a core reconfiguration on the Core Excess value. The predicted CE was compatible, within $2\sigma$, with the measured value; moreover, the net CE increase following the reconfiguration was completely satisfactory and will allow the reactor to operate at full power for several more years.\\






\section*{Bibliography}
\begin{thebibliography}{00}

\bibitem{Flussi}
	A.Borio di Tigliole, A.Cammi, D.Chiesa, M.Clemenza, S.Manera, M.Nastasi, L.Pattavina, R.Ponciroli, S.Pozzi, M.Prata, E.Previtali, 
	\emph{TRIGA reactor absolute neutron flux measurement using activated isotopes}, 
	Progress in Nuclear Energy, 
	vol.70, pp.249-255
		
\bibitem{Bayes}
	D.Chiesa, E.Previtali, M.Sisti, 
	\emph{Bayesian statistical analysis applied to NAA data for neutron flux spectrum determination}, 
	Nuclear Data Sheets, 
	vol.118, pp. 564-567
	
\bibitem{Bayes2}
	D.Chiesa, E.Previtali, M.Sisti, 
	\emph{Bayesian statistics applied to neutron activation data for reactor flux spectrum analysis}, 
	Annals of Nuclear Energy, 
	vol.70, pp. 157-168
		
\bibitem{AlCo}
	D.Chiesa, M.Clemenza, M.Nastasi, S.Pozzi, E.Previtali, G.Scionti, M.Sisti, M.Prata, A.Salvini, A.Cammi,
	\emph{Measurement and simulation of the neutron flux distribution in the TRIGA Mark~II reactor core},
	Annals of Nuclear Energy,
	vol.85, pp. 925-936
		
\bibitem{Tesi}
	D.Alloni, A.Borio di Tigliole, A.Cammi, D.Chiesa, M.Clemenza, G.Magrotti, L.Pattavina, S.Pozzi, M.Prata, E.PRevitali, A.Salvini, A.Sartori, M.Sisti, 
	\emph{Final characterization of the first critical configuration for the TRIGA Mark~II Reactor of the University of Pavia using the Monte Carlo code MCNP}, 
	Progress in Nuclear Energy, 
	vol.74, pp. 129-135
		
\bibitem{Borio}
	A.Borio di Tigliole, A.Cammi, M.Clemenza, V.Memoli, L.Pattavina, E.Previtali, 
	\emph{Benchmark evaluation of reactor critical parameters and neutron fluxes distributions at zero power for the TRIGA Mark~II reactor at the University of Pavia using the Monte Carlo code MCNP}, 
	Progress in Nuclear Energy, 
	vol.52, pp.494-502

\bibitem{CaldoPulito}
	A.Cammi, A.Zanetti, D.Chiesa, M.Clemenza, S.Pozzi, E.Previtali, M.Sisti, G.Magrotti, M.Prata,
	\emph{Characterization of the TRIGA Mark~II reactor full-power steady state},
	arXiv:1503.00873 [physics.ins-det]
	
\bibitem{Cammi}
	A.Cammi, R.Ponciroli, A.Borio di Tigliole, G.Magrotti, M.Prata, D.Chiesa, E.Previtali,
	\emph{A zero dimensional model for simulation of TRIGA Mark~II dynamic response},
	Progr. Nucl. Energy,
	vol.68, pp. 43-54
	
\bibitem{Sartori}
	A.Sartori, D.Baroli, A.Cammi, D.Chiesa, L.Luzzi, R.Ponciroli, E.Previtali, M.E.Ricotti, G.Rozza, M.Sisti,
	\emph{Comparison of a modal method and a proper orthogonal decomposition approach for multi-group time-dependent reactor spatial kinetics},
	Annals of Nuclear Energy,
	vol.71, pp. 217-229

\bibitem{MCNP}
	The X-5 Monte Carlo Team, 
	\emph{MCNP - A General Monte Carlo N-Particle Transport Code, Version 5}, 
	Los Alamos National Laboratory, 
	2005
\bibitem{Cingoli}
	A.Cambieri, F.Cingoli, S.Meloni, E.Orvini,
	\emph{Il reattore TRIGA Mark~II da 250 kV, pulsato, dell'Università di Pavia. Rapporto finale sulle prove nucleari}, 
	tech.rep., 
	University of Pavia, L.E.N.A., 
	1965




\bibitem{Stacey}
	W.M.Stacey, 
	\emph{Nuclear Reactor Physics}, 
	WILEY-VCH Verlacg GmbH \& Co. KgaA, 
	2007 




\end{thebibliography}



\end{document}